\documentclass[twocolumn,nofootinbib, superscriptaddress]{revtex4-2}
\usepackage[utf8]{inputenc}
\usepackage{natbib}
\usepackage{graphicx}
\usepackage{tikz}
\usepackage{xcolor}
\usepackage{tcolorbox}
\usepackage{hyperref}
\usepackage{cleveref}
\usepackage{float}

\usetikzlibrary{shapes.geometric, arrows.meta, positioning, fit}
\definecolor{JulienGreen}{RGB}{0, 128, 0}
\definecolor{ChristianRed}{RGB}{255, 0, 0}
\definecolor{BorisBlue}{RGB}{0, 0, 255}
\definecolor{SantiagoOrange}{RGB}{255, 165, 0}
\bibpunct{(}{)}{;}{a}{,}{,}

\usepackage{listings}
\usepackage{xcolor}
\newtcolorbox{chatgptbox}{
    colback=gray!10,
    colframe=gray!50,
    boxrule=1pt,
    arc=3mm,
    left=5mm,
    right=5mm,
    top=3mm,
    bottom=3mm,
    fonttitle=\bfseries,
    title=Freetext Output,
    before upper=\footnotesize  
}
\definecolor{codegreen}{rgb}{0,0.6,0}
\definecolor{codegray}{rgb}{0.5,0.5,0.5}
\definecolor{codepurple}{rgb}{0.58,0,0.82}
\definecolor{backcolour}{rgb}{0.95,0.95,0.92}

\lstdefinestyle{mystyle}{
    backgroundcolor=\color{backcolour},   
    commentstyle=\color{codegreen},
    keywordstyle=\color{magenta},
    numberstyle=\tiny\color{codegray},
    stringstyle=\color{codepurple},
    basicstyle=\ttfamily\footnotesize,
    breakatwhitespace=false,         
    breaklines=true,                 
    captionpos=b,                    
    keepspaces=true,                 
    numbers=left,                    
    numbersep=5pt,                  
    showspaces=false,                
    showstringspaces=false,
    showtabs=false,                  
    tabsize=2
}

\begin{document}
\title{CLAPP: The CLASS LLM Agent for Pair Programming}

\author{S.~Casas}
\affiliation{Institute for Theoretical Particle Physics and Cosmology (TTK), RWTH Aachen University, D-52056 Aachen, Germany}
\affiliation{German Aerospace Center (DLR), Scientific Information, Linder Höhe, D-51147 Köln, Germany}

\author{C.~Fidler}
\affiliation{Institute for Theoretical Particle Physics and Cosmology (TTK), RWTH Aachen University, D-52056 Aachen, Germany}

\author{B.~Bolliet}
\affiliation{Kavli Institute for Cosmology, University of Cambridge, Madingley Road, Cambridge CB3 0HA, UK}
\affiliation{Department of Physics, University of Cambridge, 19 JJ Thomson Avenue, Cambridge, CB3 0US, UK}

\author{F.~Villaescusa-Navarro}
\affiliation{Center for Computational Astrophysics, Flatiron Institute, 162 5th Ave., New York, NY 10010, USA}
\affiliation{Department of Astrophysical Sciences, Princeton University, Peyton Hall, Princeton, NJ 08544, USA}

\author{J.~Lesgourgues}
\affiliation{Institute for Theoretical Particle Physics and Cosmology (TTK), RWTH Aachen University, D-52056 Aachen, Germany}

\date{\today}
\begin{abstract}
We introduce \texttt{CLAPP} (\texttt{CLASS} LLM Agent for Pair Programming), an interactive AI assistant designed to support researchers working with the Einstein-Boltzmann solver \texttt{CLASS}. \texttt{CLAPP} leverages large language models (LLMs) and domain-specific retrieval to provide conversational coding support for \texttt{CLASS}—answering questions, generating code, debugging errors, and producing plots. Its architecture combines multi-agent LLM orchestration, semantic search across \texttt{CLASS} documentation, and a live Python execution environment. Deployed as a user-friendly web application, \texttt{CLAPP} lowers the entry barrier for scientists unfamiliar with AI tools and enables more productive human-AI collaboration in computational and numerical cosmology. The app is available at \url{https://classclapp.streamlit.app/}
\end{abstract}

\maketitle
\section{Introduction}

Modern cosmology depends critically on advanced, well-established computational tools. A prominent example is \texttt{CLASS}\footnote{\url{class-code.net}, \url{github.com/lesgourg/class_public}} \citep{class_1,class_manual,class_3} (the Cosmic Linear Anisotropy Solving System), a widely used Einstein-Boltzmann solver for calculating cosmic microwave background anisotropies and large-scale structure observables.
\texttt{CLASS} is, alongside \texttt{CAMB} \citep{Lewis:1999bs}, one of the major codes in the community and is widely used in cosmological research.
\texttt{CLASS}, like many cosmological tools, contains a large codebase with extensive documentation, which can be difficult to navigate and time-consuming to work with even for experienced researchers. For new users, the sheer size of \texttt{CLASS}—over 60,000 lines of code and 198 tunable parameters in its \texttt{explanatory.ini} configuration file—presents a significant challenge. Typically, engaging with such a codebase involves studying technical manuals, reverse-engineering examples, and extensive trial-and-error—processes that are often tedious and error-prone.

In parallel, recent advances in artificial intelligence—especially the advent of large language models (LLMs) such as \texttt{GPT-4}—have shown promise in assisting with code generation, natural language querying, and technical problem-solving. These models are increasingly capable of understanding complex instructions and producing helpful responses. However, off-the-shelf LLMs are not trained on specialized scientific software and may hallucinate incorrect details or fail to understand domain-specific context. They often also provide answers based on outdated versions of the software still found in the training data \citep{OpenAI_GPT4_2023}. They might not even be aware of newer versions of the code, as these are produced faster than the rate at which these models are retrained. Furthermore, hallucination rates in \texttt{GPT-4} remain a concern, with 28.6\% of responses being inaccurate \citep{Chelli2024}.

Recent developments in multi-agent systems (MAS) utilizing Large Language Model agents with Retrieval Augmented Generation (RAG) have shown promise in cosmological data analysis. Notable examples of the use of AI in cosmology, include:
\begin{itemize}
   \item  \textbf{CMBAgent}  \citep{Laverick:2024fyh, xu2025opensourceplanning}, which demonstrates the potential of MAS to automate and optimize research workflows in cosmology and beyond \footnote{\url{https://github.com/CMBAgents/cmbagent}}.
  \item \textbf{CosmoGemma}: A fine-tuned Gemma LLM trained on cosmology QA pairs from arXiv articles (2018–2023) that achieves 75\% accuracy and is deployed on Hugging Face to assist research productivity via question answering and idea generation. \footnote{\url{https://github.com/sultan-hassan/CosmoGemma}}.
  \item \textbf{AstroAgent}: An AI collaborator that helps astronomers generate and iteratively refine research ideas using conversation, literature review, and expert feedback, accessible via a web app and integration-ready API. \footnote{\url{https://github.com/SandyYuan/astro-agent}}.
  \item \textbf{CAMELS Agents}: AI assistants designed to help work with CAMELS (Cosmology and Astrophysics with MachinE Learning Simulations) data by answering documentation questions, searching papers, aiding coding, and drafting CAMELS sections for papers, accessible via a website requiring only LLM API keys. \footnote{\url{https://github.com/franciscovillaescusa/CAMELS_Agents}}.
  \item \textbf{ChatGaia}: An OpenAI custom-made GPT to answer astronomy questions with the Gaia Space Telescope. \footnote{\url{https://chatgpt.com/g/g-aYZOjK5zy-chatgaia}}.
  \item \textbf{AstroCoder}: AI-generated documentation for 2,270 GitHub repos from the Astrophysics Source Code Library (ASCL), providing installation instructions, examples, and a chatbot to assist with discovery and usage. \footnote{\url{https://nolank.ca/astrocoder/}}.
  \item \textbf{AstroLLama}: A specialized foundation model in astronomy. \footnote{\url{https://huggingface.co/UniverseTBD/astrollama}}.
  \item \textbf{context-assistant}: Designed to provide accurate answers by sending the full software documentation context to the AI model. The example config is for cosmology tools (CAMB, GetDist, and Cobaya), but the framework can be trivally adapted for any technical documentation.
\end{itemize}
This list is being maintained in \url{https://github.com/CMBAgents}.

\texttt{CLAPP}\footnote{\url{https://github.com/santiagocasas/clapp}} is an AI-based pair-programming assistant tailored specifically for the \texttt{CLASS} code. It provides a conversational interface that understands the structure and functionality of \texttt{CLASS} and its Python wrapper \texttt{classy}, and assists users by answering technical questions, writing and executing code, debugging, and generating plots. \texttt{CLAPP} is designed with accessibility in mind, aiming to support both experts and cosmologists with no prior experience in AI tools.

In this article, we present the design and functionality of \texttt{CLAPP}, and explain how it integrates emerging AI techniques into the computational workflows common in cosmology. We highlight three key components of its architecture: (1) multi-agent LLM orchestration, (2) retrieval-augmented generation (RAG) over a vectorized \texttt{CLASS} knowledge base, and (3) live code execution and plotting through a secure Python environment. Our goal is both pedagogical and practical: to demonstrate how LLM-based systems can be applied to real-world scientific codebases, enabling new modes of human-AI collaboration.

\section{CLAPP Architecture and Features}
\texttt{CLAPP} is built as a multi-component system that orchestrates language model agents, a knowledge database, and the \texttt{CLASS} code execution environment. At a high level, a user interacts with \texttt{CLAPP} through a chat interface, asking questions or requesting assistance with coding tasks related to \texttt{CLASS}. Under the hood, \texttt{CLAPP}'s architecture (illustrated in Fig.~1) ensures that the AI’s responses are accurate, relevant, and actionable by combining several advanced components.  We outline the main features here, corresponding to the three technical pillars of the system.

\begin{figure*}[htbp]
\centering
\includegraphics[width=\textwidth]{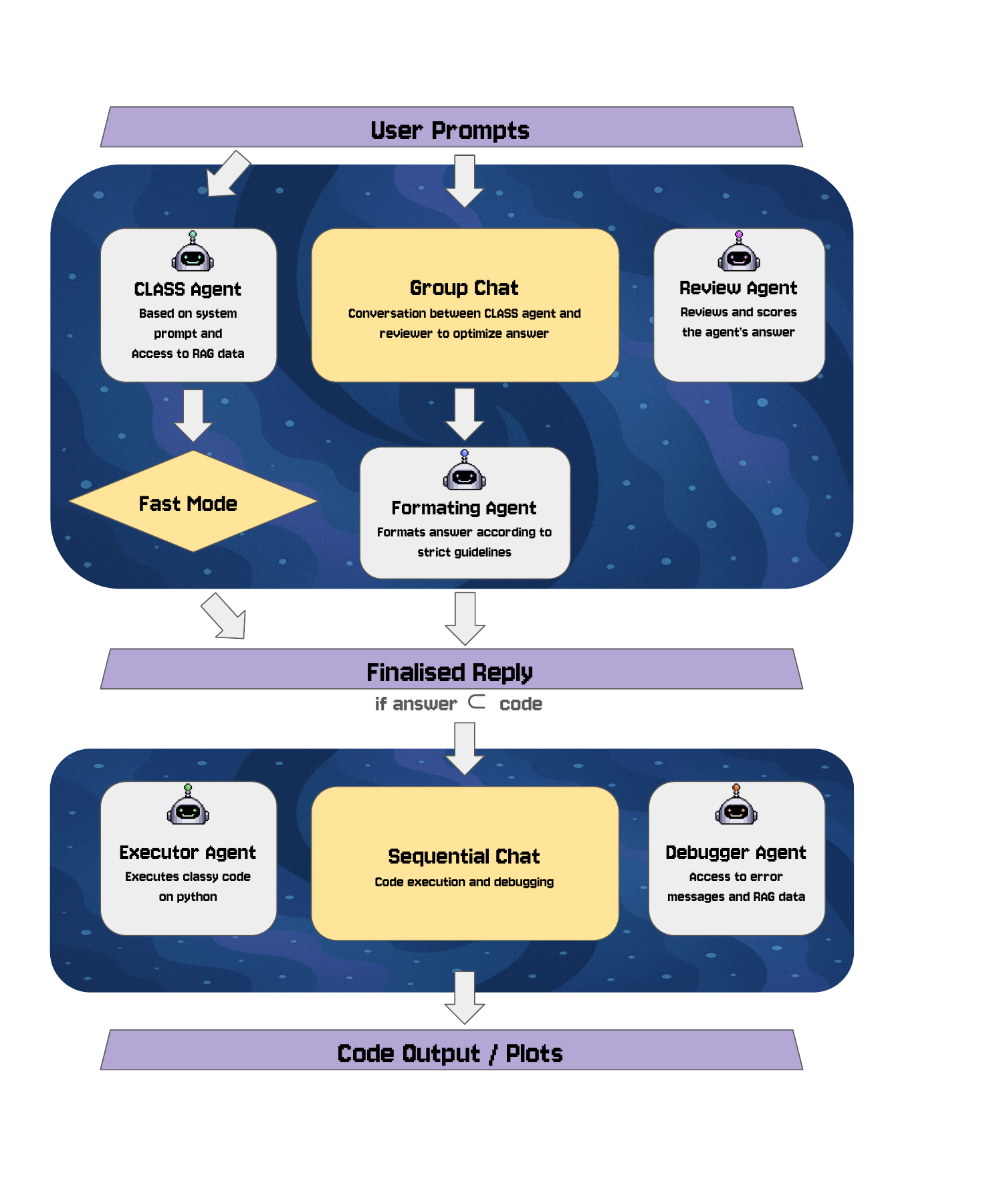}
\caption{Schematic overview of the \texttt{CLAPP} system architecture. The user interacts via a chat UI, formulating queries in natural language. The LLM agents (\texttt{CLASS} Agent, Review Agent, Formatting Agent) collaborate via a Group Chat to generate and refine answers using retrieved documentation. In Fast Mode the output of the \texttt{CLASS} agent is directly handed back to the user, as a live stream. Executable code is optionally passed to a sandbox with an Executor Agent that can call \texttt{classy} functions and a Debugger Agent that can autocorrect the code based on the error messages captured by the Executor. Results (including plots) are returned to the user. The user does not see the internal workings of \texttt{CLAPP} and only views the main results highlighted here in purple boxes. }
\label{fig:CLAPP_diagram}
\end{figure*}

\subsection{Multi-Agent LLM Orchestration}\label{sec:multiagent}

Rather than relying on a single monolithic AI agent, \texttt{CLAPP} employs a team of specialized language models that collaborate to generate high-quality responses. This architecture is built using the \emph{AG2}\footnote{\url{https://ag2.ai}} \citep{wu2023autogen, AG2_2024} multi-agent framework, which enables coordinated interaction among multiple LLMs within a shared conversational context.

In \textit{Deep Thought Mode}, a specialised \texttt{CLASS} agent first drafts a response to the user's query. A \textit{review agent} then evaluates this draft, scoring it and providing feedback on any logical, stylistic, or technical issues. Subsequently, the response is updated based on the feedback received. This review–improvement loop continues until the response meets a predefined quality threshold or a maximum number of iterations is reached. Then the final, validated response is processed by a formatting agent and returned to the user.

Each agent is powered by either an OpenAI or Gemini model and is guided by a role-specific prompt—for example, the reviewer is instructed to be especially critical and precise. This division of labor effectively simulates an automated peer-review process, allowing \texttt{CLAPP} to catch errors and refine explanations before delivering them to the user.

Alternatively, users can opt for \textit{Fast Mode}, which returns the initial draft immediately without any internal review. While this mode may be less accurate, it supports real-time streaming and is well-suited for rapid, iterative workflows.

\subsection{Retrieval-Augmented Generation with \texttt{CLASS} Knowledge}

A key feature of \texttt{CLAPP} is its ability to ground responses in the official \texttt{CLASS} documentation and codebase of the latest available \texttt{CLASS} version. To achieve this, \texttt{CLAPP} implements a Retrieval-Augmented Generation (RAG) pipeline that integrates domain-specific knowledge directly into the prompting process.

All relevant materials—including the \texttt{CLASS} user manual, parameter descriptions, example scripts, and documentation for the Python wrapper \texttt{classy}—are preprocessed and stored in a local vector database. When a user submits a query, \texttt{CLAPP} does not rely solely on the LLM’s internal knowledge (which may be incomplete or outdated). Instead, it performs a semantic search over the vectorized \texttt{CLASS} documentation to retrieve the most relevant context. This is done by embedding the user’s question into a high-dimensional vector space and matching it against precomputed embeddings of the documentation using a FAISS vector store with HuggingFace embeddings.
In the current version, we employ the embedding model \texttt{all-MiniLM-L6-v2} with a chunk size of 1000 and overlapp of 200 tokens.

The four top-matching snippets are then injected into the LLM prompt, allowing \texttt{CLAPP} to quote or paraphrase authoritative content directly from the \texttt{CLASS} materials. This approach significantly reduces hallucinations and ensures that responses are consistent with the actual behaviour and structure of the \texttt{CLASS} codebase.

For example, when asked “What does the parameter \texttt{omega\_b} do in \texttt{CLASS}?”, \texttt{CLAPP} retrieves the relevant section from the manual and accurately explains that it represents the physical baryon density—grounding the answer in the original documentation. The same mechanism applies to code generation tasks, where retrieved code snippets guide the LLM in producing correct API calls or configuration options.

The knowledge base is also modular and updateable: new content (e.g., additions to the \texttt{class-data/} directory) can be indexed and incorporated without retraining. In this way, \texttt{CLAPP} combines the generative capabilities of large language models with the reliability of curated scientific resources, producing answers that are both informative and trustworthy for cosmology researchers.

\subsection{Agent–Human Chat}

\texttt{CLAPP} is designed as an interactive chat interface between the user and the \texttt{CLAPP} AI. Internally, \texttt{CLAPP} employs a team of specialized AI agents; however, this complexity is abstracted away in the user interface, which displays only finalized, formatted responses while omitting the internal agent-to-agent dialogue.

To maintain continuity across interactions, we aggregate the outputs of both the main AI agent and the user into a conversational memory using the \emph{LangChain} framework\footnote{\url{https://www.langchain.com}} \cite{langchain2024}. This enables seamless, context-aware dialogue by preserving the history of the exchange.

In \emph{Fast Mode}, \texttt{CLAPP} additionally supports real-time message streaming, also powered by \emph{LangChain}, to enhance responsiveness.

This design allows users to engage in an iterative workflow with the AI—akin to pair programming with an expert collaborator who remembers previous steps and adapts accordingly.

\subsection{Code Execution and Plotting}

What further sets \texttt{CLAPP} apart is its integration with an actual runtime environment for \texttt{CLASS}. \texttt{CLAPP} not only talks about code – it can run code. 
Through the use of an \texttt{"Install CLASS"} button, the app downloads the full \texttt{CLASS} code, compiles it and installs its \texttt{Python} wrapper \texttt{classy} in the current \texttt{Python} environment. The user can check the installation and a plot is produced to verify that the installation was successful (see \cref{sec:AppA} for details).
When the app detects that \texttt{Python} code was generated, a \texttt{"Run Last Code Block"} button will be activated which can be pressed to execute the \texttt{classy} code. This behavior can be equally obtained by typying \texttt{"execute!"} or \texttt{"plot!"} in the agent chat.
If the code produces a plot, \texttt{CLAPP} captures the image and displays it in the chat.
If the code contains errors, the error messages are captured and a sequential chat is started where the executor and debugger agent interact with each other up to three times, to correct mistakes or bugs in the code, see lower part of \cref{fig:CLAPP_diagram}. The error message is automatically analyzed by a debugging agent, which provides a corrected version of the code – this might involve fixing a typo, adjusting a function call, or reminding the user to install missing dependencies. This error-recovery loop continues until a working solution is found or a clear explanation of the failure is provided. \texttt{CLAPP} essentially serves as a virtual “pair programmer,” catching mistakes and providing fixes in real-time. 
The ability to generate plots on the fly is particularly useful for scientific exploration: a researcher can ask for “a plot of the linear matter power spectrum for a given set of parameters,” and \texttt{CLAPP} will produce the code, execute it, and display the plot – all within the chat window. This tight integration of documentation, code generation, and execution shortens the cycle of experimentation dramatically, letting users focus on understanding results rather than wrestling with software syntax.

\section{Examples}
In this section we show some examples of inputs and outputs with \texttt{CLAPP} using different LLM models for comparison. There is no prompt optimization and both examples have been generated in Fast Mode. 

\subsection{Example: Lensed $\mathcal{C}_{\ell}$ of TT CMB}
\subsubsection*{Model: gpt-4.1}

\begin{figure}[htbp]
\centering
\includegraphics[width=0.95\linewidth]{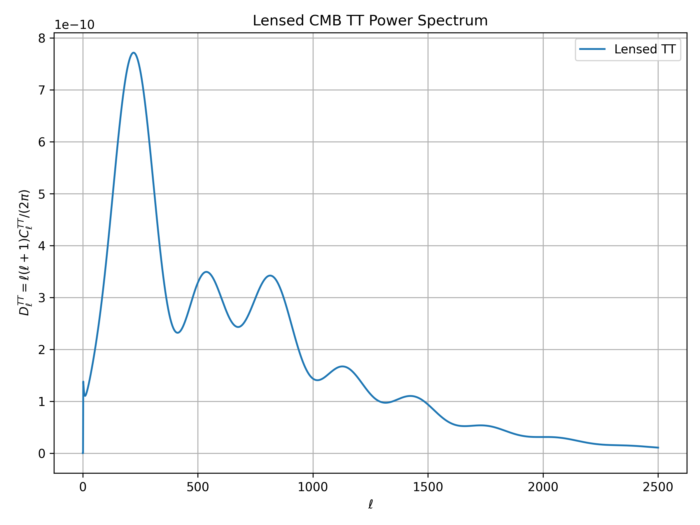}
\caption{The plot automatically generated from \texttt{CLAPP} executing the generated code for the lensed $C_{\ell}$ of CMB TT. 
The output was not manually optimised and the prompt was not tuned to generate favorable results.}
\label{fig:Cells}
\end{figure}

Example A deals with the lensed temperature power spectrum in angular space of the Cosmic Microwave Background (CMB) (see \citep{durrer2020cosmic} for a comprehensive review). The prompt is passed to the \texttt{GPT-4.1} model, one of the best models currently available from OpenAI. After executing the code, the (unaltered) plot is shown in \cref{fig:Cells}. Sometimes the code does not produce the output in the first attempt, but \texttt{CLAPP}'s self-correcting execution mode, runs up to three times until the code is error-free.

\textbf{Input:} \texttt{can you show me how to plot the lensed CMB TT C\_ells using classy?}
\textbf{Code Output}:
\begin{lstlisting}[language=Python]
from classy import Class
import numpy as np
import matplotlib.pyplot as plt

def plot_lensed_cmb_tt_cls():
    r"""
    Compute and plot the lensed CMB temperature power spectrum \( D_\ell^{TT} \) using CLASS.

    The plot shows \( D_\ell^{TT} = \ell(\ell+1)C_\ell^{TT}/(2\pi) \) as a function of multipole \( \ell \).
    All axes are dimensionless (CMB temperature anisotropy power spectrum).

    The figure is saved as 'lensed_CMB_TT_cls.png' in the current directory.
    """
    # Define cosmological parameters and CLASS options
    params = {
        'output': 'tCl,lCl',      # Request temperature Cls and lensing potential Cls
        'lensing': 'yes',         # Request lensed spectra
        'l_max_scalars': 2500,    # Maximum multipole
        'A_s': 2.1e-9,            # Scalar amplitude (dimensionless)
        'n_s': 0.965,             # Scalar spectral index (dimensionless)
        'h': 0.6732,              # Hubble parameter (dimensionless)
        'omega_b': 0.02237,       # Baryon density (dimensionless)
        'omega_cdm': 0.1200,      # CDM density (dimensionless)
        'tau_reio': 0.0544        # Reionization optical depth (dimensionless)
    }

    # Instantiate and run CLASS
    cosmo = Class()
    cosmo.set(params)
    cosmo.compute()

    # Extract the lensed Cls
    lensed_cls = cosmo.lensed_cl()  # Returns a dict with keys: 'ell', 'tt', etc.

    ell = lensed_cls['ell']
    cl_tt = lensed_cls['tt']  # Dimensionless: C_ell^TT

    # Compute D_ell = ell*(ell+1)*C_ell/(2*pi)
    D_ell = ell * (ell + 1) * cl_tt / (2 * np.pi)

    # Plot and save the spectrum
    fig, ax = plt.subplots(figsize=(8, 6))
    ax.plot(ell, D_ell, label='Lensed TT')
    ax.set_xlabel(r'$\ell$')
    ax.set_ylabel(r'$D_\ell^{TT} = \ell(\ell+1)C_\ell^{TT}/(2\pi)$')
    ax.set_title(r'Lensed CMB TT Power Spectrum')
    ax.legend()
    ax.grid(True)
    ax.relim()
    ax.autoscale_view()
    plt.tight_layout()
    plt.savefig('lensed_CMB_TT_cls.png')
    plt.close()

    print("Lensed CMB TT power spectrum plotted and saved as 'lensed_CMB_TT_cls.png'.")

    # Clean up
    cosmo.struct_cleanup()
    cosmo.empty()

plot_lensed_cmb_tt_cls()
\end{lstlisting}
\begin{chatgptbox}
    \textbf{Key Concepts:}
    Lensed CMB TT ( $C_{\ell}$ ): The angular power spectrum of the CMB temperature anisotropies, including the effect of gravitational lensing.
    classy: The Python wrapper for the CLASS code, allowing you to set cosmological parameters, run the Boltzmann solver, and extract spectra.
    \textbf{Required CLASS settings:}
        output must include 'tCl' (temperature Cls).
        lensing must be set to 'yes' to get lensed spectra.
\textbf{Step-by-Step Solution}
\begin{itemize}
    \item Set up cosmological parameters and CLASS options
You need to specify at least the basic cosmological parameters and the required outputs.
\item Run CLASS
Instantiate the Class object, set the parameters, and run the computation.
\item Extract the lensed TT ( $C_{\ell}$ ) spectrum
Use the lensed\_cl() method to get the lensed spectra.

\item Plot and save the spectrum
Use matplotlib to plot ( $D_{\ell} = \ell(\ell+1)C_{\ell}/(2\pi) )$ vs. ($\ell$), and save the plot to disk.
\end{itemize}

\textbf{Explanation}

    The code sets up a standard $\Lambda$CDM cosmology and requests lensed temperature Cls.
    The lensed\_cl() method returns a dictionary with multipoles ('ell') and spectra ('tt' for temperature).
    The plot is saved as lensed\_CMB\_TT\_cls.png in the current directory.

\textbf{Note:} You can adjust cosmological parameters as needed. The plot will not display interactively but will be saved to disk as required.

Let me know if you need help with polarization or other spectra!
\end{chatgptbox}

\subsection{Example: Suppression of the matter power spectrum from massive neutrinos}
\subsubsection*{Model: gemini-2.5-pro}

Example B, using Google's \texttt{Gemini-2.5-pro}, deals with the suppression of the power spectrum of the massive neutrinos that happens at intermediate scales, due to their free-streaming, see \citep{lesgourgues2013neutrino} for a comprehensive review. The model gives a detailed explanation of the problem and the solution that can be read in the free text output. The code produced 

\textbf{Input:} \texttt{Plot the suppression of the matter power spectrum in a massive neutrino cosmology compared to a massless case using classy}

\textbf{Code Output}:
\begin{lstlisting}[language=Python]

import matplotlib.pyplot as plt
import numpy as np
from classy import Class

# 1. Define common cosmological parameters
# These are based on Planck 2018 results, simplified for this example.
common_params = {
    'h': 0.6736,
    'omega_b': 0.02237,
    'A_s': 2.083e-9,
    'n_s': 0.9649,
    'tau_reio': 0.0544,
    'output': 'mPk',
    'P_k_max_h/Mpc': 10.0,  # Wavenumber range for P(k)
    'z_pk': 0  # Redshift for the power spectrum
}

# We fix the total matter density parameter Omega_m*h^2 to ensure the same
# late-time expansion history for a fair comparison.
# We start by defining a baseline omega_cdm for the massless case.
omega_cdm_massless = 0.1200
omega_m_h2 = omega_cdm_massless + common_params['omega_b']

# 2. Configure and run the massless neutrino model (baseline)
print("Computing massless neutrino model...")
massless_params = {
    'N_ncdm': 0,
    'N_ur': 3.044,  # Standard number of relativistic species
    'omega_cdm': omega_cdm_massless
}
massless_params.update(common_params)

massless_cosmo = Class()
massless_cosmo.set(massless_params)
massless_cosmo.compute()

# 3. Configure and run the massive neutrino model
print("Computing massive neutrino model...")
# We'll use a total mass of M_nu = 0.1 eV.
M_nu = 0.1  # Total neutrino mass in eV

# Calculate the physical density of massive neutrinos, omega_ncdm.
# The conversion factor is M_nu = 93.14 * omega_ncdm.
omega_ncdm = M_nu / 93.14

# Adjust omega_cdm to keep total matter density constant.
omega_cdm_massive = omega_m_h2 - common_params['omega_b'] - omega_ncdm

# For 1 massive species, to keep N_eff ~ 3.044, we set N_ur = 2.0308.
massive_params = {
    'N_ncdm': 1,
    'm_ncdm': M_nu,
    'N_ur': 2.0308,
    'omega_cdm': omega_cdm_massive
}
massive_params.update(common_params)

massive_cosmo = Class()
massive_cosmo.set(massive_params)
massive_cosmo.compute()

# 4. Extract power spectra and calculate the suppression ratio
# Define a k-range in h/Mpc.
k_vals = np.logspace(-4, 1, 500)
h = common_params['h']

# Get P(k) from CLASS. classy returns P(k) in (Mpc/h)^3 units
# if k is given in h/Mpc.
pk_massless = np.array([massless_cosmo.pk(k * h, 0) for k in k_vals]) * h**3
pk_massive = np.array([massive_cosmo.pk(k * h, 0) for k in k_vals]) * h**3

# Calculate the ratio.
pk_ratio = pk_massive / pk_massless

# 5. Generate the plot
print("Generating plot...")
plt.style.use('classic')
fig, (ax1, ax2) = plt.subplots(
    2, 1, sharex=True, gridspec_kw={'hspace': 0.05}, figsize=(8, 7))
fig.suptitle(r'Effect of Massive Neutrinos on the Matter Power Spectrum', fontsize=16)

# Top panel: Plot the absolute matter power spectra
ax1.loglog(k_vals, pk_massless, label=r'Massless $\nu$ ($M_\nu=0$ eV)')
ax1.loglog(k_vals, pk_massive, linestyle='--', label=r'Massive $\nu$ ($M_\nu=0.1$ eV)')
ax1.set_ylabel(r'$P(k) \quad [(\mathrm{Mpc}/h)^3]$', fontsize=14)
ax1.grid(True)
ax1.legend(loc='lower left')
ax1.set_title(r'Comparison at $z=0$ with fixed $\Omega_m h^2$')

# Bottom panel: Plot the suppression ratio
ax2.semilogx(k_vals, pk_ratio)
ax2.axhline(y=1, color='gray', linestyle='--')
ax2.set_xlabel(r'$k \quad [h/\mathrm{Mpc}]$', fontsize=14)
ax2.set_ylabel(r'$P_{\mathrm{massive}} / P_{\mathrm{massless}}$', fontsize=14)
ax2.grid(True)
ax2.set_ylim(0.8, 1.02)  # Zoom in on the suppression effect

# Save the figure to disk
plt.savefig('matter_power_spectrum_neutrino_suppression.png', dpi=300, bbox_inches='tight')
print("Plot saved as 'matter_power_spectrum_neutrino_suppression.png'")

# Clean up CLASS instances
massless_cosmo.struct_cleanup()
massless_cosmo.empty()
massive_cosmo.struct_cleanup()
massive_cosmo.empty()
\end{lstlisting}

\begin{chatgptbox}
Of course. Massive neutrinos, being a form of hot dark matter, suppress the growth of cosmic structures on small scales. This leads to a distinct suppression in the matter power spectrum, $P(k)$, at high wavenumbers ($k$). We can visualize this effect by comparing a standard cosmological model with massless neutrinos to one with massive neutrinos, keeping other key parameters constant.
\begin{enumerate}
    \item Key Concepts
    
     \textbf{Matter Power Spectrum, P(k):} This is the Fourier transform of the matter two-point correlation function. It quantifies the amount of structure (density fluctuations) at a given comoving scale, represented by the wavenumber k. Large k corresponds to small scales, and small k corresponds to large scales.
     
     \textbf{Massive Neutrinos (ncdm):} In CLASS, non-cold dark matter (ncdm) components, like massive neutrinos, are treated separately from cold dark matter (cdm). Unlike CDM, massive neutrinos have significant thermal velocities in the early universe.
     Neutrino Free-Streaming: Due to their high thermal velocities, massive neutrinos can travel large distances, effectively "streaming" out of small gravitational potential wells. This erases density perturbations on scales smaller than their free-streaming length, thus slowing down the overall growth of structure.
     \textbf{Cosmological Parameters:} To isolate the effect of neutrino mass, we will compare two models. The baseline model will have only massless neutrinos (N\_ncdm=0). The second model will have a massive neutrino component (N\_ncdm=1). We will keep the total matter density parameter, Omega\_m h\^2, constant between the two models. This is a standard approach that ensures the late-time expansion history is the same, allowing for a direct comparison of structure growth. We achieve this by reducing the cold dark matter density (omega\_cdm) to compensate for the mass added by the neutrinos (omega\_ncdm).

    \item Step-by-Step Solution

     \textbf{Define a Baseline (Massless) Model:} We will first configure a standard $\Lambda$CDM cosmology with purely massless neutrinos (N\_ncdm = 0, N\_ur = 3.044). We fix cosmological parameters like h, omega\_b, A\_s, and n\_s to values consistent with recent observations (e.g., Planck 2018).
     \textbf{Define a Massive Neutrino Model:} We will then create a second model with one massive neutrino species with a total mass M\_nu = 0.1 eV. To keep the total matter density (omega\_m = omega\_cdm + omega\_b + omega\_ncdm) constant, we will calculate the energy density of the massive neutrinos (omega\_ncdm) and subtract it from the omega\_cdm of the baseline model.
     \textbf{Compute Power Spectra:} Using classy, we will compute the matter power spectrum, P(k), for both models at redshift z=0.
     \textbf{Calculate the Suppression Ratio:} We will calculate the ratio of the massive neutrino P(k) to the massless neutrino P(k). This ratio will be close to 1 on large scales (small k) and will drop below 1 on small scales (large k), clearly showing the suppression.
     \textbf{Plot the Results:} We will generate a two-panel plot. The top panel will show both power spectra, and the bottom panel will show their ratio to highlight the scale-dependent suppression.
    \end{enumerate}
\end{chatgptbox}
\begin{figure}[htbp]
\centering
\includegraphics[width=0.95\linewidth]{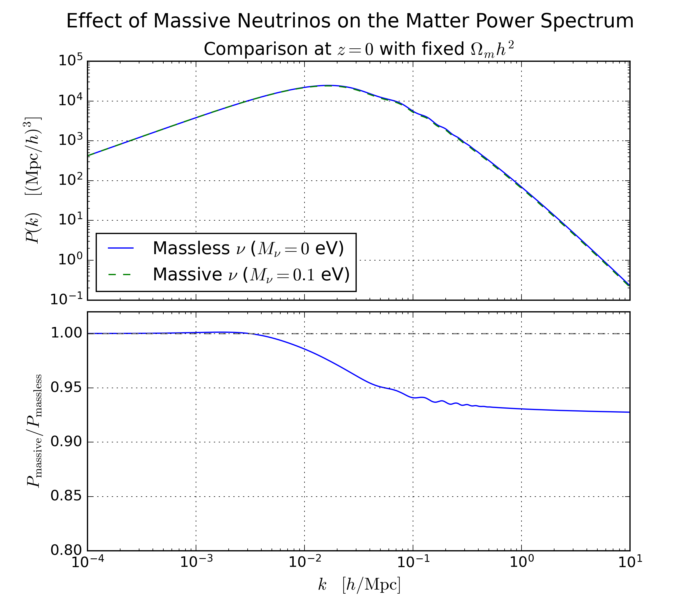}
\caption{The plot automatically generated from \texttt{CLAPP} executing the generated code for the neutrino case in Example B. The upper panel shows the linear matter power spectrum, while the lower panel shows the ratio of the massive to the massless case, showcasing the suppression of power happening at intermediate scales.}
\label{fig:Neutrinos}
\end{figure}

\section{Deployment and Use Cases}
\texttt{CLAPP} is implemented as a web application using \emph{Streamlit}\footnote{https://streamlit.io/} a web-based service to deploy apps in Python. This means researchers can start using \texttt{CLAPP} through their web browser without any complex installation, effectively bringing the power of AI-assisted coding to anyone with an internet connection. On first use, the user is prompted to provide an API key for the LLM (since \texttt{CLAPP} leverages the OpenAI API for GPT-4, or Google's Gemini models). The user is also able to store their keys safely on the Streamlit cloud using encryption. Once initialized, the interface presents a chat window and a few control options. Users unfamiliar with LLMs do not need to know any special syntax – they can type questions like “How do I modify the initial conditions in \texttt{CLASS} for a non-standard cosmology?” or requests like “Write a Python snippet to compute the matter power spectrum using \texttt{CLASS}.” The system will handle the rest: retrieving documentation about initial conditions or using known \texttt{CLASS} Python API patterns to draft the code. 

Throughout the interaction, \texttt{CLAPP} strives to maintain a user-friendly conversational tone, explaining steps in plain language while also providing technical details and references to \texttt{CLASS} documentation where appropriate. When it executes code, it clearly delineates the code and the results. 
\texttt{CLAPP} is an ideal application for the casual \texttt{CLASS} user who wants to get meaningful results from this extensive Einstein-Boltzmann solver and its Python wrapper \texttt{classy} without delving into its intricate details. Ultimately, for the advanced user it also provides information about the \texttt{CLASS} C-code documentation and how to modify its source code. However, in this paper we have not focused too much on this use case, since validating the replies from the agents in that case is rather tedious.

\subsection{Conclusion}
\texttt{CLAPP} represents a step toward integrating AI pair-programming tools into the daily practice of a cosmologist. By combining a state-of-the-art language model with domain-specific retrieval and a live execution environment, \texttt{CLAPP} functions as an intelligent assistant tuned for cosmology workflows. Importantly, we have focused on usability for researchers who are not AI experts: the system is accessible with a simple web interface, but it encapsulates powerful capabilities behind the scenes. Early experiments with \texttt{CLAPP} suggest that even users with limited \texttt{CLASS} experience can perform complex tasks with its guidance, and advanced users can accelerate their work by delegating routine tasks to the AI. Looking forward, this approach could be extended to other scientific codes and domains—imagine similar LLM-based assistants for $N$-body simulations, astrophysical data analysis, or instrument pipelines. As large language models continue to improve, and as we refine multi-agent coordination and tool integration, we anticipate that AI assistants like \texttt{CLAPP} will become increasingly valuable collaborators in research, not replacing human scientists but augmenting their capabilities. In cosmology, where bridging theory, code, and data is often the hard part, \texttt{CLAPP} offers a glimpse of a future where those bridges are easier to cross with the help of AI. \vspace{1em}

\textit{Source code and demonstration:} The \texttt{CLAPP} project is open-source, with code available on GitHub (\url{https://github.com/santiagocasas/clapp}). We invite the cosmology community to experiment with \texttt{CLAPP}, provide feedback, and contribute to its development.

\bibliographystyle{apsrmp} 

\subsection*{Acknowledgments}
The work of BB was partially funded by an unrestricted gift from Google, the Cambridge Centre for Data-Driven Discovery Accelerate Programme and the Infosys-Cambridge AI Centre.
We would like to thank Nils Schöneberg for enhancing the \texttt{CLASS} documentation, which led to considerable improvements in the RAG data and, consequently, the performance of \texttt{CLASS}. Our gratitude also extends to other current and former members of the TTK at RWTH Aachen University for their initial testing of the app.
\subsubsection*{Author Contributions}
\textbf{SC and CF}: idea, implementation of agents, back-end and front-end code development, streamlit app, testing, validating, main writing.
\textbf{BB}: agents, langchain and ag2 implementation, ideas, testing, writing.
\textbf{FVN}: initial ideas, langchain support, testing, writing.
\textbf{JL}: testing, documentation, validation, writing.

\appendix

\section{Detailed usage instructions}\label{sec:AppA}

In this appendix we provide very detailed instructions of the CLAPP code for beginners.
Note that CLAPP can be used in two ways:
\begin{itemize}
\item online on \url{https://classclapp.streamlit.app/}, for easy and immediate use;
\item on individual computers, after cloning the code on \url{https://github.com/santiagocasas/clapp} and following the installation instructions in the \texttt{readme}.
\end{itemize}
The latter can be useful if the user wants to customize the app: for isntance, one may enhance the knowledge of CLAPP on specific aspects by adding more documentation or notebooks to the RAG. In this Appendix, we focus on the use of the online version.

\begin{figure}[htbp]
    \centering
    \includegraphics[width=0.95\linewidth]{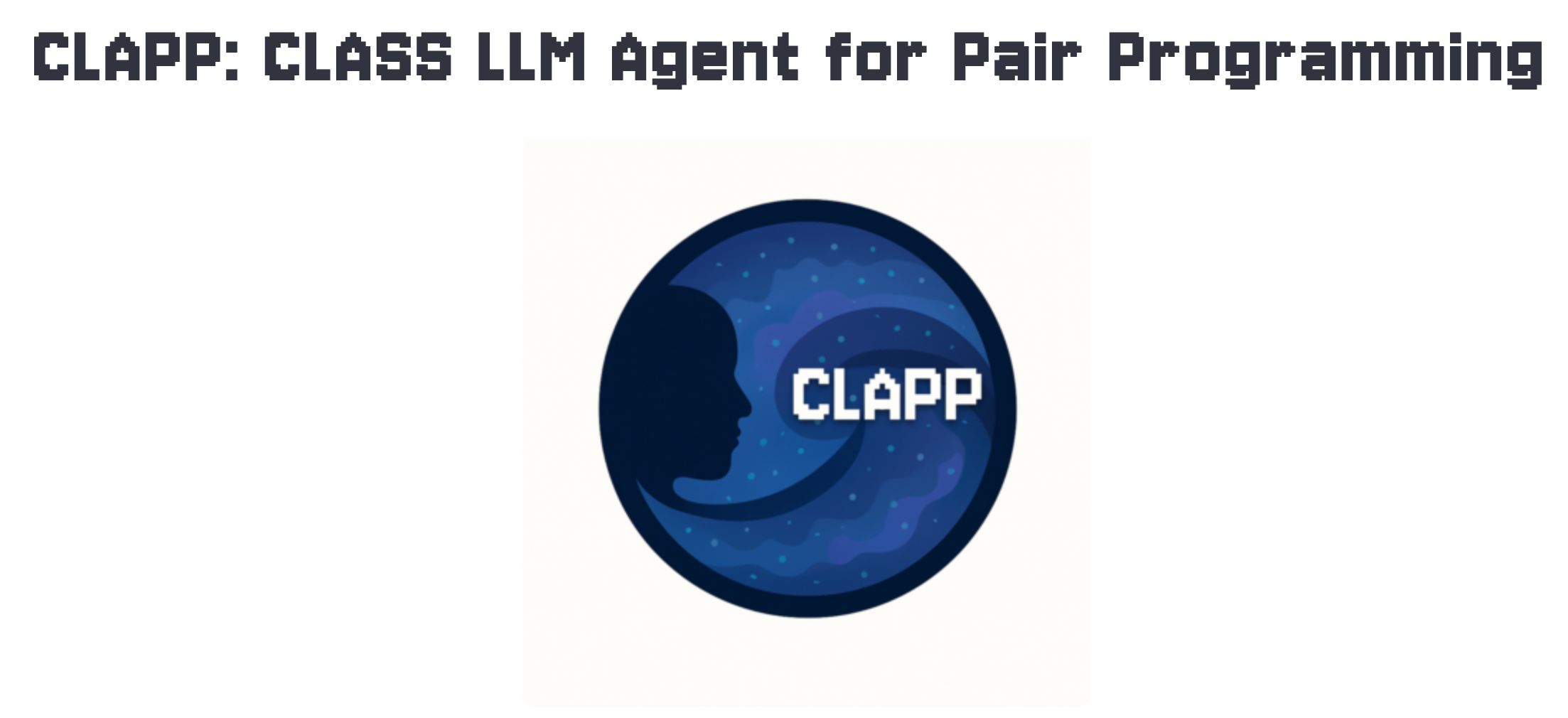}
    \caption{CLAPP: CLASS LLM Agent for Pair Programming. Logo of the app.}
    \label{fig:enter-label}
\end{figure}

\subsection*{API keys}
For the app to work, the most important thing is to set up the API keys. If the user does not have an OpenAI paid subscription, they can find free-to-use API keys for the Google Gemini models at: \url{https://aistudio.google.com/apikey}.

\begin{figure}[htbp]
    \centering
    \includegraphics[width=0.55\linewidth]{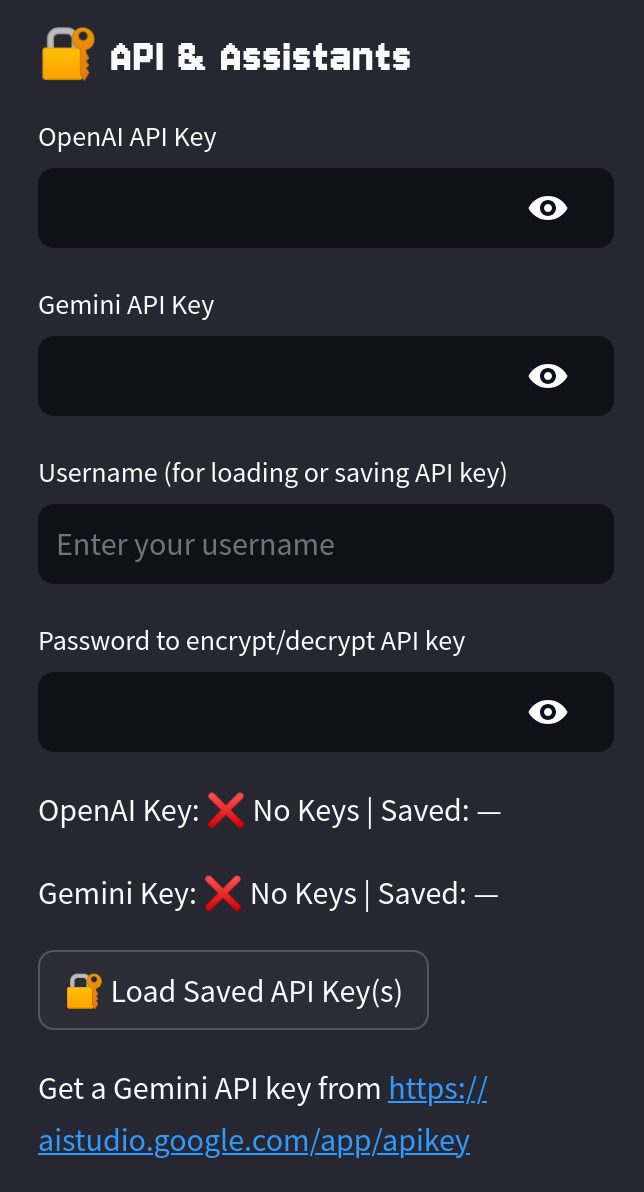}
    \caption{Interface to introduce the API keys of your favorite LLM provider. Username and password are optional and just for easy access to your personal keys. We do not take reponsibility for lost or misused keys.}
    \label{fig:keys}
\end{figure}

The API keys can be stored on the Streamlit cloud in a secure way, by typing a username and a password. The keys are then encrypted and stored, so that the user does not need to remember the long API keys from the LLM provider.
See \cref{fig:keys} for a preview of the app.

\subsection*{Choosing an LLM model}
It is important to choose the correct LLM model for the given task. The available models for OpenAI are: 
\texttt{GPT\_MODELS = "gpt-4o-mini", "gpt-4o", "gpt-4.1"}
while for Gemini they are:
\texttt{GEMINI\_MODELS = "gemini-2.5-flash", "gemini-2.5-flash-lite", "gemini-2.5-pro"}.

\begin{figure}[htbp]
    \centering
    \includegraphics[width=0.55\linewidth]{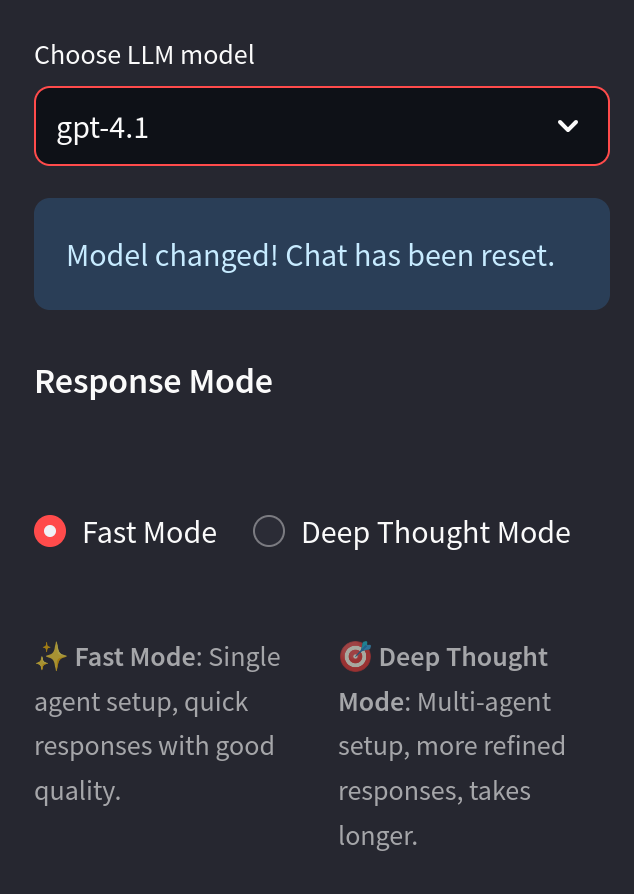}
    \caption{Available models, once Gemini's and OpenAI's API keys are inserted.}
    \label{fig:models}
\end{figure}

In \cref{fig:models} the drop-down menu for the different models can be seen, as well as the radio buttons for \textit{Fast Mode} and \textit{Deep Thought Mode} explained in the text in \cref{sec:multiagent}.

We recommend \texttt{"gpt-4o-mini"} and \texttt{"gemini-2.5-flash-lite"} for quick, simple testing, while for more advanced tasks, especially in \textit{Deep Thought Mode} we recommend \texttt{"gpt-4.1"} and \texttt{"gemini-2.5-flash"}. The \texttt{"gemini-2.5-pro"} model has its own reasoning mode, so it could be much slower when used in our multi-agent setup. The latter might change in the near future, according to Google Gemini's free usage policies.

\subsection*{Further settings}

\begin{figure}[h]
    \centering
    \includegraphics[width=0.55\linewidth]{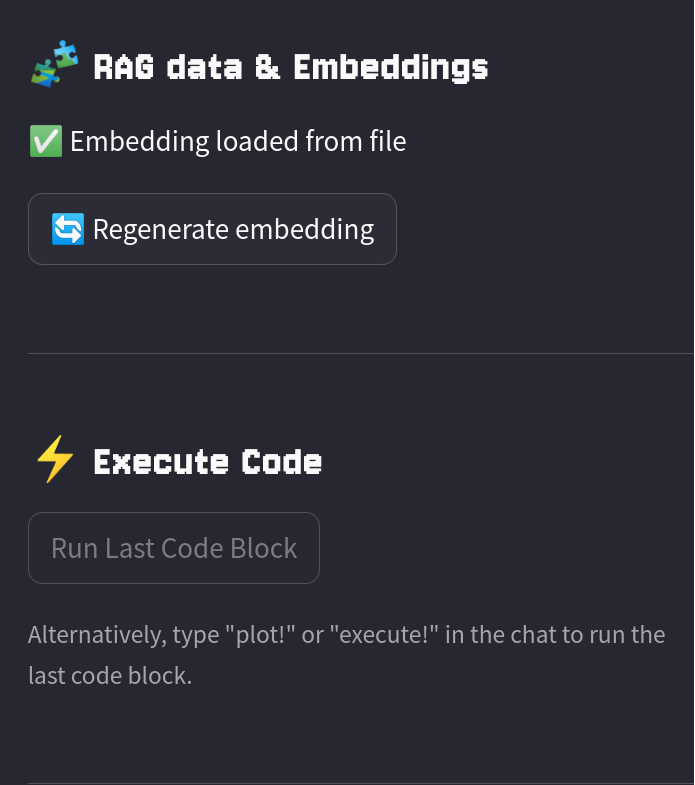}
    \\
    \includegraphics[width=0.55\linewidth]{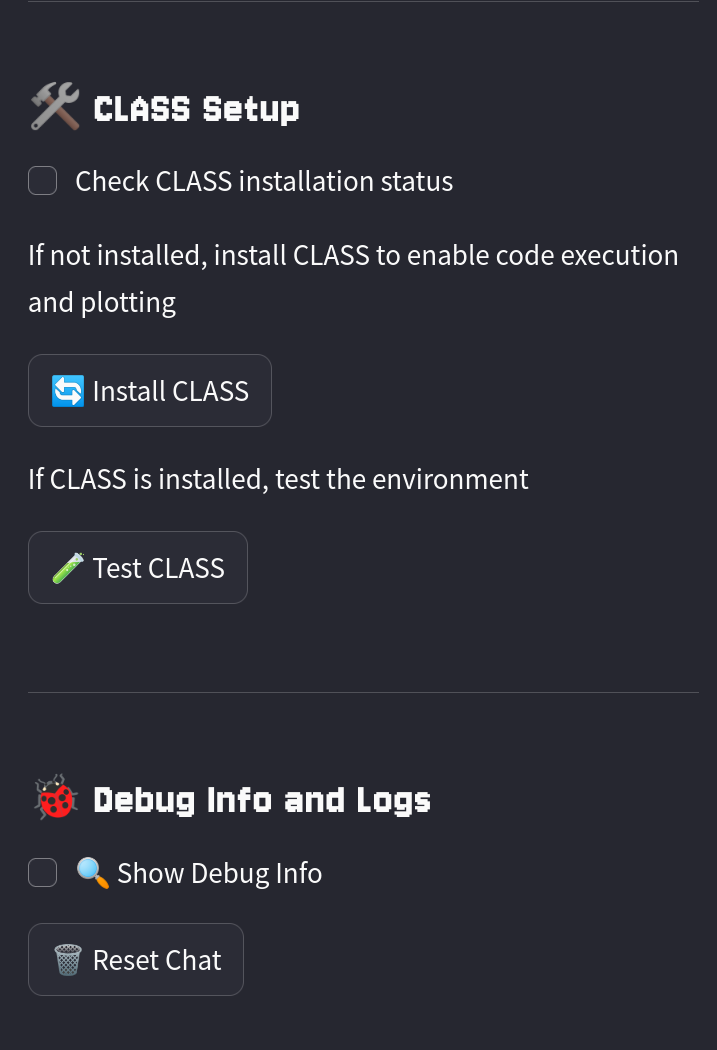}
    \caption{Further settings for RAG data embedding, \texttt{CLASS} installation and code execution.
    \textbf{Top:} The button for "Regenerate embedding" should only be used in local mode, if the RAG data has been modified or enhanced. If this is pressed on the \textit{Streamlit} app, the result will be the same as before, while the embedding will take a few minutes to be completed. The button "Run Last Code Block" should be used whenever some Python code has been generated by the \texttt{CLAPP} agent. Alternatively, the user can type "plot!" or "execute!" on the chat.
    \textbf{Bottom:} With the checkbox, the user can check for the \texttt{CLASS} installation setup. If a positive result is obtained, the button "Test CLASS" runs a test script that should produce a plot visible on the sidebar. If the installation check is negative, the user can install \texttt{CLASS} with the "Install CLASS" button. This should download \texttt{CLASS} and install \texttt{classy} into the Python execution environment.}
    \label{fig:controls}
\end{figure}

In \cref{fig:controls} we show other settings available in the app. 

\textbf{RAG data \& Embeddings:}
The RAG embedding explained in the main text is already shipped with the Streamlit app on the cloud. Therefore, this button should not be executed unless you have changed the RAG data. This can happen if you run your Streamlit app locally. See \url{https://docs.streamlit.io/} for instructions and our GitHub repo (\url{https://github.com/santiagocasas/clapp}) for local installation instructions.

\textbf{CLASS Setup:}
Ideally, the Streamlit app will ship with \texttt{CLASS} installed by default. However, if the Streamlit app gets rebooted from scratch, the installation has to be performed only once. Here the user can first test the \texttt{CLASS} installation with a checkbox and if the code is not installed, then by pressing "Install CLASS" the \texttt{CLASS} code will be cloned, compiled and its Python wrapper \texttt{classy} will be installed to the Python environment.
The installation can be tested with the "Test CLASS" button and the user should be able to see a plot in the sidebar.

\textbf{Debug Info and Logs:}
With this setting the user will be able to see some app debug info from the backend and also the context being passed to the agents from the RAG. We do not recommend this for the standard user.
Finally, the "Reset Chat" button is helpful if the web chat has already been pupulated with code, plots or maybe too many error messages.
As usual, it is important to test \texttt{CLAPP} thoroughly by playing with it and doing several tests of trial-and-error.
\clearpage
\bibliography{clappbib}

\begin{thebibliography}{13}
\expandafter\ifx\csname natexlab\endcsname\relax\def\natexlab#1{#1}\fi
\expandafter\ifx\csname bibnamefont\endcsname\relax
  \def\bibnamefont#1{#1}\fi
\expandafter\ifx\csname bibfnamefont\endcsname\relax
  \def\bibfnamefont#1{#1}\fi
\expandafter\ifx\csname citenamefont\endcsname\relax
  \def\citenamefont#1{#1}\fi
\expandafter\ifx\csname url\endcsname\relax
  \def\url#1{\texttt{#1}}\fi
\expandafter\ifx\csname urlprefix\endcsname\relax\def\urlprefix{URL }\fi
\providecommand{\bibinfo}[2]{#2}
\providecommand{\eprint}[2][]{\url{#2}}

\bibitem[{\citenamefont{Blas} \emph{et~al.}(2011)\citenamefont{Blas, Lesgourgues, and Tram}}]{class_manual}
\bibinfo{author}{\bibnamefont{Blas}, \bibfnamefont{D.}}, \bibinfo{author}{\bibfnamefont{J.}~\bibnamefont{Lesgourgues}}, and \bibinfo{author}{\bibfnamefont{T.}~\bibnamefont{Tram}}, \bibinfo{year}{2011}, \bibinfo{journal}{JCAP} \textbf{\bibinfo{volume}{07}}, \bibinfo{pages}{034}.

\bibitem[{\citenamefont{Chelli} \emph{et~al.}(2024)\citenamefont{Chelli, Descamps, Lavou{\'{e}}, Trojani, Azar, Deckert, Raynier, Clowez, Boileau, and Ruetsch-Chelli}}]{Chelli2024}
\bibinfo{author}{\bibnamefont{Chelli}, \bibfnamefont{M.}}, \bibinfo{author}{\bibfnamefont{J.}~\bibnamefont{Descamps}}, \bibinfo{author}{\bibfnamefont{V.}~\bibnamefont{Lavou{\'{e}}}}, \bibinfo{author}{\bibfnamefont{C.}~\bibnamefont{Trojani}}, \bibinfo{author}{\bibfnamefont{M.}~\bibnamefont{Azar}}, \bibinfo{author}{\bibfnamefont{M.}~\bibnamefont{Deckert}}, \bibinfo{author}{\bibfnamefont{J.-L.} \bibnamefont{Raynier}}, \bibinfo{author}{\bibfnamefont{G.}~\bibnamefont{Clowez}}, \bibinfo{author}{\bibfnamefont{P.}~\bibnamefont{Boileau}}, and \bibinfo{author}{\bibfnamefont{C.}~\bibnamefont{Ruetsch-Chelli}}, \bibinfo{year}{2024}, \bibinfo{journal}{J Med Internet Res} \textbf{\bibinfo{volume}{26}}(\bibinfo{number}{1}), \bibinfo{pages}{e53164}, \urlprefix\url{https://www.jmir.org/2024/1/e53164/}.

\bibitem[{\citenamefont{Durrer}(2020)}]{durrer2020cosmic}
\bibinfo{author}{\bibnamefont{Durrer}, \bibfnamefont{R.}}, \bibinfo{year}{2020}, \emph{\bibinfo{title}{The cosmic microwave background}} (\bibinfo{publisher}{Cambridge University Press}).

\bibitem[{\citenamefont{{LangChain}}(2024)}]{langchain2024}
\bibinfo{author}{\bibnamefont{{LangChain}}}, \bibinfo{year}{2024}, \bibinfo{title}{Langchain documentation}, \bibinfo{howpublished}{\url{https://python.langchain.com/docs/}}, \bibinfo{note}{accessed: 2025-07-09}.

\bibitem[{\citenamefont{Laverick} \emph{et~al.}(2024)\citenamefont{Laverick, Surrao, Zubeldia, Bolliet, Cranmer, Lewis, Sherwin, and Lesgourgues}}]{Laverick:2024fyh}
\bibinfo{author}{\bibnamefont{Laverick}, \bibfnamefont{A.}}, \bibinfo{author}{\bibfnamefont{K.}~\bibnamefont{Surrao}}, \bibinfo{author}{\bibfnamefont{I.}~\bibnamefont{Zubeldia}}, \bibinfo{author}{\bibfnamefont{B.}~\bibnamefont{Bolliet}}, \bibinfo{author}{\bibfnamefont{M.}~\bibnamefont{Cranmer}}, \bibinfo{author}{\bibfnamefont{A.}~\bibnamefont{Lewis}}, \bibinfo{author}{\bibfnamefont{B.}~\bibnamefont{Sherwin}}, and \bibinfo{author}{\bibfnamefont{J.}~\bibnamefont{Lesgourgues}}, \bibinfo{year}{2024}, \bibinfo{title}{Multi-agent system for cosmological parameter analysis}, \eprint{2412.00431}, \urlprefix\url{https://arxiv.org/abs/2412.00431}.

\bibitem[{\citenamefont{Lesgourgues}(2011)}]{class_1}
\bibinfo{author}{\bibnamefont{Lesgourgues}, \bibfnamefont{J.}}, \bibinfo{year}{2011}, \bibinfo{note}{1093 citations counted in INSPIRE as of 23 May 2025}, \eprint{1104.2932}.

\bibitem[{\citenamefont{Lesgourgues} \emph{et~al.}(2013)\citenamefont{Lesgourgues, Mangano, Miele, and Pastor}}]{lesgourgues2013neutrino}
\bibinfo{author}{\bibnamefont{Lesgourgues}, \bibfnamefont{J.}}, \bibinfo{author}{\bibfnamefont{G.}~\bibnamefont{Mangano}}, \bibinfo{author}{\bibfnamefont{G.}~\bibnamefont{Miele}}, and \bibinfo{author}{\bibfnamefont{S.}~\bibnamefont{Pastor}}, \bibinfo{year}{2013}, \emph{\bibinfo{title}{Neutrino cosmology}} (\bibinfo{publisher}{Cambridge University Press}).

\bibitem[{\citenamefont{Lesgourgues and Tram}(2011)}]{class_3}
\bibinfo{author}{\bibnamefont{Lesgourgues}, \bibfnamefont{J.}}, and \bibinfo{author}{\bibfnamefont{T.}~\bibnamefont{Tram}}, \bibinfo{year}{2011}, \bibinfo{journal}{JCAP} \textbf{\bibinfo{volume}{09}}, \bibinfo{pages}{032}, \bibinfo{note}{311 citations counted in INSPIRE as of 23 May 2025}.

\bibitem[{\citenamefont{Lewis} \emph{et~al.}(2000)\citenamefont{Lewis, Challinor, and Lasenby}}]{Lewis:1999bs}
\bibinfo{author}{\bibnamefont{Lewis}, \bibfnamefont{A.}}, \bibinfo{author}{\bibfnamefont{A.}~\bibnamefont{Challinor}}, and \bibinfo{author}{\bibfnamefont{A.}~\bibnamefont{Lasenby}}, \bibinfo{year}{2000}, \bibinfo{journal}{Astrophys. J.} \textbf{\bibinfo{volume}{538}}, \bibinfo{pages}{473}.

\bibitem[{\citenamefont{OpenAI}(2023)}]{OpenAI_GPT4_2023}
\bibinfo{author}{\bibnamefont{OpenAI}}, \bibinfo{year}{2023}, \bibinfo{journal}{arXiv preprint arXiv:2303.08774} \urlprefix\url{https://arxiv.org/abs/2303.08774}.

\bibitem[{\citenamefont{Wang} \emph{et~al.}(2024)\citenamefont{Wang, Wu, and the AG2~Community}}]{AG2_2024}
\bibinfo{author}{\bibnamefont{Wang}, \bibfnamefont{C.}}, \bibinfo{author}{\bibfnamefont{Q.}~\bibnamefont{Wu}}, and \bibinfo{author}{\bibnamefont{the AG2~Community}}, \bibinfo{year}{2024}, \bibinfo{title}{Ag2: Open-source agentos for ai agents}, \bibinfo{note}{available at https://docs.ag2.ai/}, \urlprefix\url{https://github.com/ag2ai/ag2}.

\bibitem[{\citenamefont{Wu} \emph{et~al.}(2023)\citenamefont{Wu, Bansal, Zhang, Wu, Li, Zhu, Jiang, Zhang, Zhang, Liu} \emph{et~al.}}]{wu2023autogen}
\bibinfo{author}{\bibnamefont{Wu}, \bibfnamefont{Q.}}, \bibinfo{author}{\bibfnamefont{G.}~\bibnamefont{Bansal}}, \bibinfo{author}{\bibfnamefont{J.}~\bibnamefont{Zhang}}, \bibinfo{author}{\bibfnamefont{Y.}~\bibnamefont{Wu}}, \bibinfo{author}{\bibfnamefont{B.}~\bibnamefont{Li}}, \bibinfo{author}{\bibfnamefont{E.}~\bibnamefont{Zhu}}, \bibinfo{author}{\bibfnamefont{L.}~\bibnamefont{Jiang}}, \bibinfo{author}{\bibfnamefont{X.}~\bibnamefont{Zhang}}, \bibinfo{author}{\bibfnamefont{S.}~\bibnamefont{Zhang}}, \bibinfo{author}{\bibfnamefont{J.}~\bibnamefont{Liu}}, \emph{et~al.}, \bibinfo{year}{2023}, \bibinfo{journal}{arXiv preprint arXiv:2308.08155} .

\bibitem[{\citenamefont{Xu} \emph{et~al.}(2025)\citenamefont{Xu, Sarkar, Lonappan, Íñigo Zubeldia, Villanueva-Domingo, Casas, Fidler, Amancharla, Tiwari, Bayer, Ekioui, Cranmer} \emph{et~al.}}]{xu2025opensourceplanning}
\bibinfo{author}{\bibnamefont{Xu}, \bibfnamefont{L.}}, \bibinfo{author}{\bibfnamefont{M.}~\bibnamefont{Sarkar}}, \bibinfo{author}{\bibfnamefont{A.~I.} \bibnamefont{Lonappan}}, \bibinfo{author}{\bibnamefont{Íñigo Zubeldia}}, \bibinfo{author}{\bibfnamefont{P.}~\bibnamefont{Villanueva-Domingo}}, \bibinfo{author}{\bibfnamefont{S.}~\bibnamefont{Casas}}, \bibinfo{author}{\bibfnamefont{C.}~\bibnamefont{Fidler}}, \bibinfo{author}{\bibfnamefont{C.}~\bibnamefont{Amancharla}}, \bibinfo{author}{\bibfnamefont{U.}~\bibnamefont{Tiwari}}, \bibinfo{author}{\bibfnamefont{A.}~\bibnamefont{Bayer}}, \bibinfo{author}{\bibfnamefont{C.~A.} \bibnamefont{Ekioui}}, \bibinfo{author}{\bibfnamefont{M.}~\bibnamefont{Cranmer}}, \emph{et~al.}, \bibinfo{year}{2025}, \bibinfo{title}{Open source planning and control system with language agents for autonomous scientific discovery}, \eprint{2507.07257}, \urlprefix\url{https://arxiv.org/abs/2507.07257}.

\end{thebibliography}
\end{document}